\documentclass[fleqn,usenatbib]{mnras}

\usepackage{newtxtext,newtxmath}

\usepackage[T1]{fontenc}

\DeclareRobustCommand{\VAN}[3]{#2}
\let\VANthebibliography\thebibliography
\def\thebibliography{\DeclareRobustCommand{\VAN}[3]{##3}\VANthebibliography}

\usepackage{graphicx}	
\usepackage{amsmath}	

\newcommand{\Gam}{$\Gamma$}
\newcommand{\zai}{$\xi$}

\newcommand{\aref}[1]{\hyperref[#1]{Appendix~\ref{#1}}}

\defcitealias{Krumholz+17}{KTOM}

\title[Molecular galactic winds]{When are galactic winds molecular?}

\author[Vijayan \& Krumholz]{
Aditi Vijayan$^{1,2}$\thanks{E-mail: aditi.vijayan@anu.edu.au (AV)}
and Mark R. Krumholz$^{1,2}$
\\
$^{1}$Research School of Astronomy and Astrophysics, Australian National University, Cotter Road, Weston ACT 2612, Australia\\
$^{2}$ARC Centre of Excellence for Astronomy in Three Dimensions (ASTRO-3D), Canberra ACT 2601, Australia
}

\date{Accepted XXX. Received YYY; in original form ZZZ}

\pubyear{2015}

\begin{document}
\label{firstpage}
\pagerange{\pageref{firstpage}--\pageref{lastpage}}
\maketitle

\begin{abstract}
The molecular phase of supernova-driven outflows originates from the cold, molecular gas in the disc of a star-forming galaxy, and may carry a substantial fraction of the wind mass flux in some galaxies, but it remains poorly understood. Observations of this phase come mostly from very nearby galaxies due its low surface brightness and covering fraction, and simulations often lack the spatial resolution necessary to resolve it. Here we analytically estimate the survivability of this phase in order to understand under what conditions an galactic wind can contain a significant molecular phase. We show that the molecular content of outflows is primarily determined by two dimensionless numbers: a generalised Eddington ratio describing the strength of the outflow and an ionisation parameter-like quantity describing the strength of the radiation field per baryon. We apply this model to a sample of galaxies and show that, while any molecules entrained in the winds of normal star-forming galaxies should be destroyed close to the galactic disc, the outflows of strong starburst should become increasingly dominated by molecules.
\end{abstract}




\section{Introduction}
\label{sec:intro}

In star-forming galaxies, supernova-driven multiphase, panchromatic outflows act as a conduit for mass, momentum, energy and metal exchange between a galaxy and its surrounding gas. Theoretical work suggests that different phases of the outflows are responsible for carrying different parts of this flow, with hot gas dominating the energy and metal flux while cooler gas dominates the mass flux \citep[e.g.,][]{Li20a, Kim+20, Kim20b, Schneider20a, Rathjen21a, Vijayan24a}. This suggests that understanding the cool phase of the outflows is key for understanding the broader baryon cycle as well as the regulation of star formation within galaxies. However, ``cool'' gas -- usually denoting gas at temperatures $\lesssim 10^4$ K -- is not a single phase. Instead, it can be further subdivided into a phase at $T\sim 10^4$ K where the hydrogen is mostly ionised, a neutral hydrogen phase with temperatures $T \sim 200 - 7000$ K, and a molecular phase where most hydrogen is locked into H$_2$ molecules and the gas temperature falls to $T \sim 10 - 100$ K. In galaxies with star formation rates high enough to drive substantial outflows, particularly beyond $z = 0$, the molecular phase usually dominates the mass budget of the interstellar medium \citep{Tacconi20a}, but it is unclear to what extent this translates to the mass budget of the outflows.

The question of how cool outflows further subdivide into ionised, neutral, and molecular components has proven difficult to study in numerical simulations due to resolution. Even the transition between the $T\sim 10^6$ K and $T\sim 10^4$ K ionised phases is challenging to capture properly in simulations because it depends on processes such as Kelvin-Helmholtz and Rayleigh-Taylor instabilities in turbulent boundary layers that are seldom to never resolved in galaxy-scale simulations \citep[e.g.,][and references therein]{Fielding22a}. The neutral and molecular phases, which are typically much denser than the $\sim 10^4$ K ionised phase, are even harder to resolve, and consequently many simulations do not even attempt to include the cooling or chemical processes relevant to them \citep[though there are exceptions, e.g.,][]{Kim20b, Rathjen21a, Chen23a, Vijayan24a}. Similarly, semi-analytic models for multiphase outflows \citep{Huang20b, Huang22a, Weinberger23a, Smith24a, Butsky24b}, which are calibrated from these simulations, generally ignore the distinction between molecular, atomic, and cool ionised gas, instead lumping them into a single ``cool'' phase.

The phase structure of the cool gas has also proven difficult to study observationally. While the ionised component is seen ubiquitously in both absorption and emission studies \citep[][and references therein]{Veilleux20a}, the atomic and molecular phases are much more challenging to study because their small filling factor makes them difficult to see in absorption, and their low surface brightness makes them difficult to see in emission. To date there have been only a few detections of these components of outflows from star-forming galaxies reported in literature. Perhaps the best-studied example is M82, where both atomic \citep{Martini18a} and molecular \citep{Leroy15b} gas have been mapped from the base of the outflow where it joins the ISM up to distances of $\sim$ few kpc above the disc. Quantitative analysis of these data indicates that the atomic and molecular phases likely carry as much or more mass flux than the cool ionised phase \citep{Yuan23a}. Other local examples of molecular outflows include NGC 253 \citep{Bolatto+13, Walter+17} and the Large Magellanic Cloud \citep{Tchernyshyov22a}, and there is indirect evidence from stacks of high redshift galaxies that many also posses large molecular outflows \citep{Ginolfi17a}. Finally, in the central region of the Milky Way, \citet{Di-Teodoro18a, Di-Teodoro20a} detect a large population of outflowing clouds containing a mix of neutral and molecular hydrogen. \citet{Noon23a} show that these clouds are predominantly molecular close to the Galactic plane but become increasingly atomic at large distances, and that they are characterised by a disequilibrium chemical state suggesting that the molecules are in the process of photodissociating as the clouds move away from the Galaxy.

Given the difficulty of studying in particular the atomic and molecular parts of outflows from both simulations and observations, it is helpful to turn to analytic models that can be used as a guide for future work. That is our goal here: in this work, we present a predictive theoretical model for quantifying the molecular fraction in a galactic outflow, focusing in particular on the balance between the neutral and ionised phases, the two that are least-studied by both observations and theory. We explore the parameter space of galaxy masses, star formation rates, and dynamical times and understand which of these are important for a galaxy to host molecular outflows. Our ultimate goal is a first-order answer to the basic question: for which galaxies is it likely that a substantial component of the mass flux will be carried by the molecular phase of the outflow? The layout of the paper is as follows. \autoref{sec:the_model} we introduce the model and its assumptions. In \autoref{sec:results} we apply the model to make predictions about the molecular content of outflows as a function of galaxy properties, and apply those predictions to observed galaxies. We conclude and discuss future prospects in \autoref{sec:conclusion}.

\section{The Model}\label{sec:the_model}

Here we provide a model to understand the atomic-molecular balance in outflows, and in particular to understand under what circumstances we expect the non-ionised part of the outflow to contain a substantial molecular component, versus when we expect it to be fully atomic. We begin in \autoref{ssec:rad_vs_coll} with a general discussion of radiative versus collisional processes, and then in \autoref{ssec:model} we use this discussion to develop a simple but predictive model for the atomic-to-molecular ratio in single cloud, which we extend to the full wind in \autoref{ssec:total_flux}.

\subsection{Radiative versus collisional processes}
\label{ssec:rad_vs_coll}

As noted in \autoref{sec:intro}, the bulk of the mass available to be launched into outflows in most galaxies with substantial star formation rates exists in molecular form in the disc, and spatially resolved observations in both the Milky Way centre and M82 indicate that near their launch point most of the mass in outflows is molecular, with the balance shifting toward atomic further from the galaxy. However, free atomic hydrogen is a higher energy state than hydrogen molecules, and in the absence of a dissociative process a cloud of free atomic hydrogen will spontaneously convert to H$_2$ on timescales of Myr. Thus if an outflow contains significant amounts of atomic hydrogen, there must be some mechanism that dissociates H$_2$ molecules and inhibits this conversion. This mechanism could be either collisional -- for example, mixing of supernova-heated gas into the outflow which keeps it warm enough to dissociate H$_2$ -- or radiative, in the form of Lyman-Werner band photons that excite the H$_2$ molecules into the $B^1\Sigma_u$ or $C^1\Pi_u$ electronic states, from which they can de-excite into the unbound continuum of the ground $X^1\Sigma_g^+$ electronic state. Our first task in constructing a model for molecular outflow is to determine which of these channels is likely to be dominant.

To answer this question we can compare collisional and radiative dissociation rates. In a predominantly atomic medium the main collisional dissociation channel is
\begin{equation}
    \mathrm{H}_2 + \mathrm{H} \to 3\mathrm{H},
\end{equation}
so the collisional dissociation rate per unit time per H$_2$ molecule is $k_\mathrm{H} n_\mathrm{H}$; we take our rate coefficient $k_\mathrm{H}$ as a function of $n_\mathrm{H}$ and gas temperature $T$ from equation 26 of \citet{Glover07a}, and note that, as expected, $k_\mathrm{H}$ is a monotonically increasing function of $T$. The competing photodissociation channel occurs at a rate that depends on the intensity of the radiation field. The \citet{Draine78a} field that is typical of the Milky Way near the Solar neighbourhood produces a dissociation rate $D_0 = 5.8\times 10^{-11}$ s$^{-1}$ \citep{Sternberg14a}, so we take the photodissociation rate to be $\chi D_0$, with $\chi=1$ corresponding to the \citeauthor{Draine78a} field. Thus the condition $n_\mathrm{H} k_\mathrm{H} = \chi D_0$ implicitly specifies a minimum temperature $T_\mathrm{coll}$ above which the collisional channel dominates and below which the radiative channel does.

However, the temperature cannot be arbitrarily high if H~\textsc{i} is to be present -- if $T$ is too large, then the gas will collisionally ionise and become dominated by H$^+$ instead. We define $T_\mathrm{ion}$ as the temperature below which gas in collisional ionisation equilibrium will have $<50\%$ of its hydrogen ionised. We compute $T_\mathrm{ion}$ as a function of $n_\mathrm{H}$ using the astrochemistry code \textsc{despotic} \citep{Krumholz14b}, using the \citet{Gong17a} chemical network with an ambient non-ionising radiation field $\chi = 1$, zero ionising radiation, and a cosmic ray ionisation rate of $2\times 10^{-16}$ s$^{-1}$ \citep{Indriolo12a}. We see that the hydrogen reaches 50\% collisional ionisation at $T\approx 1.6 \times 10^4$ K, with only an extremely weak dependence on density.

\begin{figure}
    \includegraphics[width=\columnwidth]{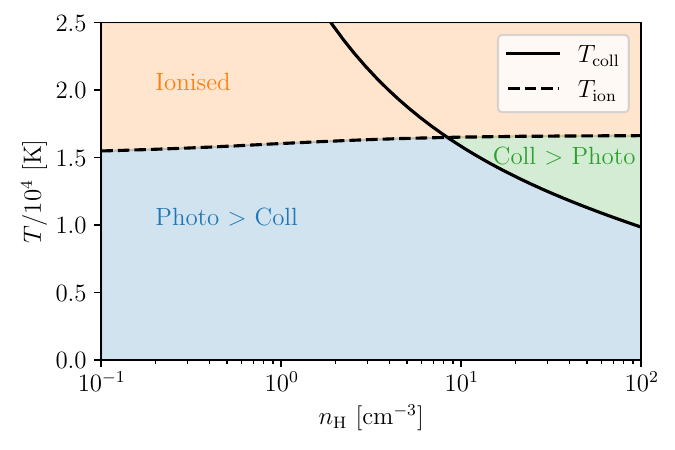}
    \caption{The temperatures $T_\mathrm{coll}$ at which collisional dissociation becomes more important that photodissociation (solid line), and $T_\mathrm{ion}$ at which hydrogen is 50\% collisionally ionised (dashed line), as a function of number density of H nuclei $n_\mathrm{H}$. These lines divide the plane into three regions: one where photodissociation is the dominant producer of atomic hydrogen (blue), one where collisional dissociation dominates (green), and one where no neutral gas is present due to collisional ionisation (orange). We see that collisional dissociation can only ever be dominant at densities $n_\mathrm{H} \gtrsim 10$ cm$^{-3}$, and then only over a narrow range of temperatures from $\approx 1.0 - 1.6\times 10^4$ K.
    \label{fig:coll_vs_rad}
    }
\end{figure}

Thus the condition for dissociation to be dominant over photoionisation reduces to the requirement that $T_\mathrm{coll} < T < T_\mathrm{ion}$. We plot $T_\mathrm{coll}$ and $T_\mathrm{ion}$ as a function of $n_\mathrm{H}$, for $\chi = 1$, in \autoref{fig:coll_vs_rad}. The plot shows that collisional dissociation is never more important than photodissociation at densities $\lesssim 10$ cm$^{-3}$, and that at higher densities collisional dissociation dominates only over a fairly narrow range in temperature, from $\approx 1.0 - 1.6\times 10^4$ K. Our conclusion based on this is that photodissociation will be dominant in most circumstances, and we will therefore proceed to build our model for the chemical state of outflows focusing on this process.

\subsection{The atomic-to-molecular transition in a single cloud}
\label{ssec:model}

\begin{figure}
    	\includegraphics[width=\columnwidth]{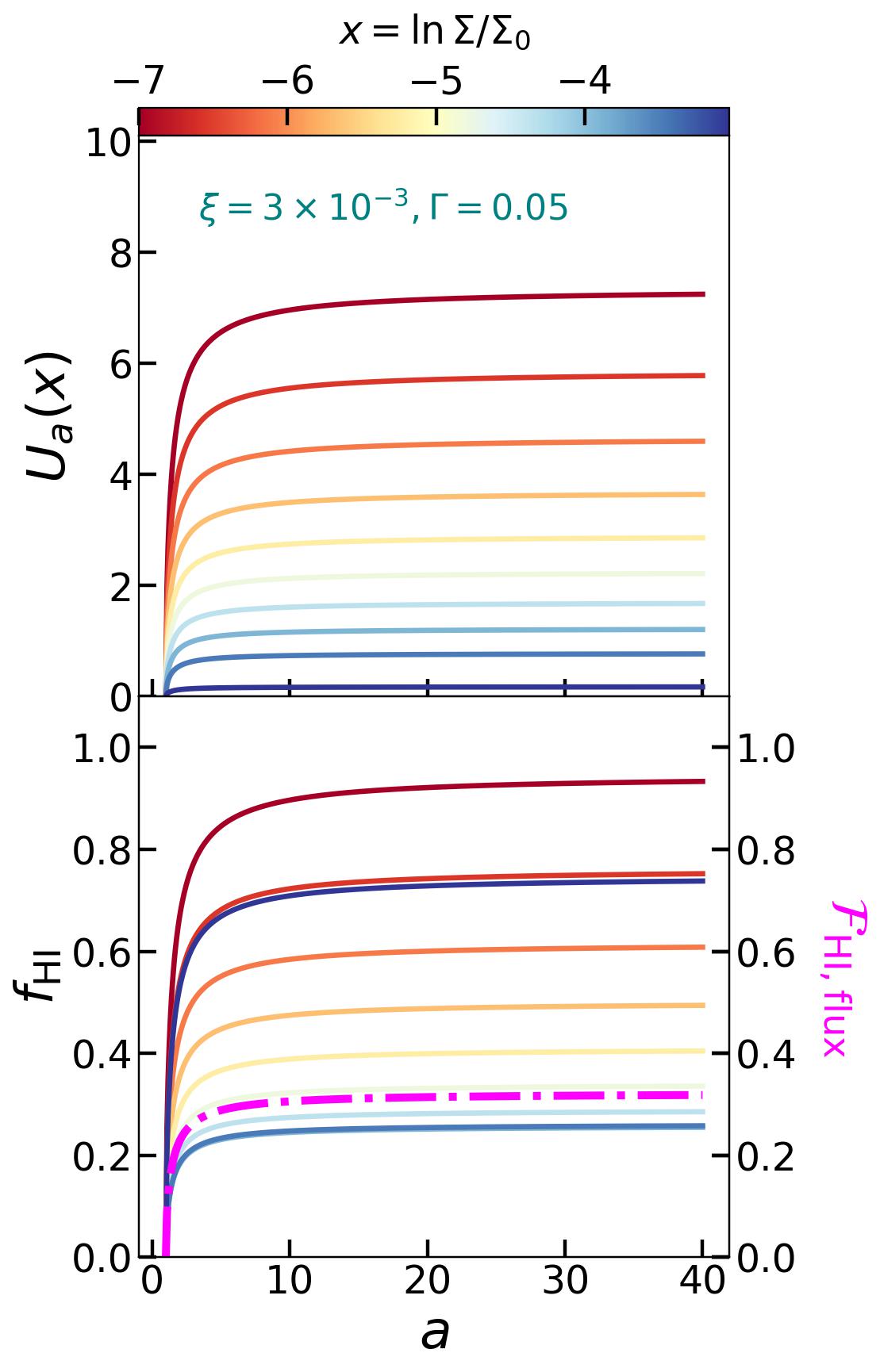}
    \caption{The acceleration law (top), H~\textsc{i} fraction $f_\mathrm{HI}$ (bottom, solid curves), and flux-averaged H~\textsc{i} fraction $\mathcal{F}_\mathrm{HI,flux}$ (bottom, dot-dashed magenta line), for a wind with clouds of fixed area driven by a galaxy with a point source potential; different coloured solid lines indicate different values of the dimensionless cloud surface density $x$, as indicated in the colour bar. The example shown uses the parameters $\xi=3\times 10^{-3}$, $\Gamma=0.05$, and $\mathcal{M}=1000$.}
    \label{fig:uax_fhi}

\end{figure}

Now that we have determined that our focus should be on radiative rather than collisional processes, we build a model to determine when radiation is expected to yield outflow with a high molecular content, versus when it is expected to efficiently dissociate most H$_2$ molecules, which we build on top of the model for momentum-driven cool winds proposed by \citet{Krumholz+17}, building on earlier work by \citet{Thompson16a}, to which we refer hereafter as the \citetalias{Krumholz+17} model. This model has been validated by comparison to both full 3D simulations \citep{Raskutti17a} and observations of the wind of M82 \citep{Yuan23a}. In this model, clouds are continuously launched at a radius $r_0$ from a star-forming disc with mean surface density $\overline{\Sigma}_0$. The clouds in the disk possess a log-normal distribution of surface densities. A cloud will be launched into the wind if its surface density, $\Sigma$, satisfies $x=\ln \Sigma/\overline{\Sigma}_0<x_{\rm crit}=\ln \Gamma$, where $\Gamma$ is the generalised Eddington ratio, quantifying the competition between gravity and momentum injection -- $\Gamma = 1$ corresponds to a galaxy where feedback and gravitational forces are equal for a cloud with surface density $\overline{\Sigma}_0$. 

Once launched, a cloud moves radially outward, following a trajectory whereby its velocity at any given radius can be written $v(r) = v_0 U_a(x)$, where $v_0 = \sqrt{2 G M_0 / r_0}$ is the escape speed from $r_0$, $M_0$ is the gravitating mass interior to $r_0$, and $a = r / r_0$ is a dimensionless radius. The function $U_a(x)$ is called the wind acceleration law, and in the idealised case of a wind driving mechanism that simply supplies momentum at a fixed rate per unit area -- the only case we will consider here -- depends on the initial dimensionless cloud surface density $x$, on the generalised Eddington factor for the wind \Gam, on the shape of the gravitational potential (e.g., point mass-like versus isothermal), and on the rate at which clouds' cross-sectional areas expand as they travel outward in the wind. We refer readers for \citetalias{Krumholz+17} for a discussion of how $U_a(x)$ is calculated, and for the remainder of this paper simply adopt their results.

\subsubsection{Fixed Area Case}\label{ssec:fixed_area}


We first consider the simplest case of clouds that maintain a fixed cross-sectional area as they travel outward. For a point-like gravitational potential, these clouds obey an acceleration law,
\begin{equation}\label{eqn:uax}
    U_a^2 (x) = \left(\Gamma e^{-x} - 1 \right) \left(\frac{a-1}{a}\right)\,.
\end{equation}
For the purpose of illustration, in the top panel of \autoref{fig:uax_fhi} we plot $U_a(x)$ as a function of $a$ for a few different values of $x$. For this fixed-area, point-potential case, the clouds accelerate rapidly up to $a\lesssim 5$ and then move with a constant speed.

Now let us consider a cloud that is entirely molecular at the time when it enters the wind, consistent with observations showing that outflows are largely molecular at their base, and estimate how quickly it will be converted to a purely atomic one. As we have seen the dominant process will be photodissociation, and a radiation field of normalised strength $\chi$ provides a dissociating photon flux $F^* = 2\times 10^7 \chi$ photons cm$^2$ s$^{-1}$ on the surface of a molecular cloud \citep{Krumholz+09}. For small cloud column densities ($\lesssim 10$ M$_\odot$ pc$^{-2}$ at Solar metallicity, and thus applicable to most clouds with column densities small enough to be picked up by the wind), this radiation field will result in an H$_2$ dissociation rate that greatly exceeds the formation rate, so the mass per unit time per unit area dissociated from molecular to atomic composition will be $\dot{\Sigma}_\mathrm{diss} = 2 m_\mathrm{H} f_\mathrm{diss} F^* \equiv \chi \dot{\Sigma}_\mathrm{diss,0}$, where $\dot{\Sigma}_\mathrm{diss,0}= 1.5$ M$_\odot$ pc$^{-2}$ Myr$^{-1}$ is the dissociation rate expected for $\chi = 1$, and $f_\mathrm{diss} \approx 0.15$ is the fraction of photon absorptions that result in a dissociation rather than radiative de-excitation back to a bound molecular state \citep{Noon23a}. We let $\chi_0$ be the radiation field intensity at $r_0$, and assume that the radiation field is dominated by the central galaxy, and therefore falls off as $\chi \propto 1/r^2$. With this assumption, we can write the rate of change in atomic hydrogen surface density as $\dot{\Sigma}_{\mathrm{HI}} = \chi_0 \dot{\Sigma}_\mathrm{diss,0}  / a^2$, and the rate of change with respect to radius is therefore
\begin{equation}
    \frac{d\Sigma_\mathrm{HI}}{da} =
    \frac{\chi_0 t_0 \dot{\Sigma}_\mathrm{diss,0}}{a^2 U_a(x)}, 
\end{equation}
for $\Sigma_\mathrm{HI} < \Sigma$; here we have defined $t_0 = r_0/v_0$.

We are interested in tracking the atomic mass fraction, $f_{\rm HI}=\Sigma_{\rm HI}/{\overline{\Sigma}_0} e^{x}$, as a function of distance $a$. We can express this as 
\begin{equation}
    \frac{d f_{\rm HI}}{da} =  \frac{\chi_0\dot{\Sigma}_\mathrm{diss,0} t_0}{\overline{\Sigma}_0 e^x} \frac{1}{a^2 U_a(x)}\,,
\end{equation}
for $f_\mathrm{HI} < 1$. The above equation has an exact solution with the boundary condition $f_{\rm HI} (a=1)=0$,
\begin{equation}\label{eqn:fhi}
    f_{\rm HI} = \min\left(\frac{2 \chi_0 \dot{\Sigma}_\mathrm{diss,0} t_0 e^{-x}}{\overline{\Sigma}_0 \sqrt{\Gamma e^{-x}-1}} \sqrt{\frac{a-1}{a}}, 1\right)\,.
\end{equation}
We plot $f_{\rm HI}$ as a function of $a$ in the middle panel of \autoref{fig:uax_fhi} for a range of $x$ values and using the sample parameters $\Gamma=0.05$ and $\xi = 3\times 10^{-3}$. As with $U_a(x)$, we see that $f_{\rm HI}$ attains a steady value at $a\gtrsim5$. Clouds with larger surface density at the beginning of their trajectory reach atomic fractions of $\approx 25\%$, while lower surface density clouds become increasingly atomic-dominated.

\subsubsection{Variable Area Case}\label{subsec:var_area}

The expression we have just derived applies to clouds that maintain a constant cross-sectional area as they travel outward in the wind. However, \citetalias{Krumholz+17} also consider the possibility of clouds that expand as they travel outward, with a cloud radius that expands $\propto a^2$ (constant solid angle) or $\propto a$ (intermediate between constant solid angle and constant area). In general if the radius of a cloud varies with distance as $a^p$ for some power $p$, then the column when the cloud is at distance $a$ is $\Sigma = \overline{\Sigma}_0 e^x a^{-p}$, and therefore evolves as $d\Sigma/da = -p\Sigma/a$. If we assume that the atomic and molecular parts expand homologously, then it is clear that the evolution equation for the atomic surface density generalises to
\begin{equation}
    \frac{d\Sigma_\mathrm{HI}}{da} = \frac{\chi_0 t_0 \dot{\Sigma}_\mathrm{diss,0}}{a^2 U_a(x)} - p\frac{\Sigma_\mathrm{HI}}{a}
\end{equation}
and the evolution equation for the atomic fraction is
\begin{equation}
    \frac{df_\mathrm{HI}}{da} = \frac{d}{da}\left(\frac{\Sigma_\mathrm{HI}}{\Sigma}\right) = \frac{\chi_0\dot{\Sigma}_\mathrm{diss,0} t_0}{\overline{\Sigma}_0 e^x} \frac{a^{p-2}}{U_a(x)}
\end{equation}
This clearly reduces to the constant area case we have derived above when $p=0$. We can therefore express the H~\textsc{i} fraction at some distance $a$ for a cloud characterised by dimensionless (log) surface density $x$ as
\begin{equation}
    f_\mathrm{HI}(a,x) = \min\left(\xi e^{-x} \int_1^a \frac{a'^{p-2}}{U_{a'}(x)} \, da', 1\right),
    \label{eq:fHI}
\end{equation}
where we define
\begin{equation}
    \xi \equiv \frac{\chi_0\dot{\Sigma}_\mathrm{diss,0} t_0}{\overline{\Sigma}_0}.
    \label{eq:xi_def}
\end{equation}
The integral appearing in \autoref{eq:fHI} cannot be evaluated analytically for all possible wind acceleration laws as it can be for the case of constant area and a point gravitational potential, but it is straightforward to evaluate numerically. The dimensionless quantity $\xi$ in \autoref{eq:fHI} can be thought of as analogous to an ionisation parameter, in that it describes a ratio of the radiation flux to the baryon density, or as describing a ratio of timescales: $t_0$ is the characteristic time that it takes a cloud that is launched in the wind to move away from the galaxy, while $\overline{\Sigma}_0/\chi_0 \dot{\Sigma}_\mathrm{diss,0}$ is the characteristic timescale required to dissociate a cloud of molecular hydrogen with a surface density equal to the mean surface density of the galaxy, $\overline{\Sigma}_0$. Values of $\xi \gg 1$ correspond to galaxies where dissociation happens fast compared to motion, values $\xi \ll 1$ to cases where motion is faster.

\subsection{The total and asymptotic molecular flux}
\label{ssec:total_flux}

The above expression applies to a single cloud with a fixed initial (log) surface density $x$; we can use this to calculate the total molecular fraction for a system in which clouds at a range of column densities are continuously being ejected from the galaxy. In the \citetalias{Krumholz+17} model, the PDF of surface density with respect to mass follows a log-normal form, which we reproduce here from Equation 8 of \citetalias{Krumholz+17}:
\begin{equation}
    p_M(x) = \frac{1}{\sqrt{2 \pi \sigma_x^2}} \rm{exp} \Big[-\frac{(x-\sigma_x^2/2)^2}{2 \sigma_x^2} \Big]\,,
\end{equation}
where here $\sigma_x$ is the dispersion of the log-normal, which depends on the Mach number $\mathcal{M}$ of the turbulence in the galaxy. Clouds are ejected only for $x < x_\mathrm{crit}$.

Given this distribution, the total wind mass flux carried by clouds with initial column densities in the rate $x$ to $x+dx$ is $d\dot{M} \propto p_M(x) \, dx$, and the atomic mass flux is $d\dot{M}_\mathrm{HI} \propto f_\mathrm{HI}(a,x) p_M(x) \, dx$, where $f_\mathrm{HI}$ is given by \autoref{eq:fHI}. We are therefore now in a position to write down the flux-averaged atomic fraction in the wind as
\begin{equation}
    \mathcal{F}_\mathrm{HI}(a) = \frac{\int_{-\infty}^{x_\mathrm{crit}} f_\mathrm{HI}(a,x) p_M(x) \, dx}{\int_{-\infty}^{x_\mathrm{crit}} p_M(x) \, dx}.
    \label{eq:Fhi_flux}
\end{equation}
We define $\mathcal{F}_\mathrm{HI} \equiv \lim_{a\to\infty} \mathcal{F}_\mathrm{HI}(a)$ as the asymptotic atomic flux, i.e., the fraction of the flux that reaches infinity in the form of atomic hydrogen; conversely, $\mathcal{F}_\mathrm{H_2} = 1 - \mathcal{F}_\mathrm{HI}$ is the fraction that reaches infinity in molecular form.

The dot-dashed magenta line in the bottom panel of \autoref{fig:uax_fhi} shows $\mathcal{F}_\mathrm{HI}(a)$ estimated for same values of $\Gamma$ and $\xi$ mentioned in \autoref{ssec:fixed_area}, and using a Mach number $\mathcal{M} = 1000$, a choice that we will justify below. Even though there are individual clouds that may be entirely atomic at large $a$, the flux-averaged fraction of atomic gas is $\sim 0.3$. Following the trends in $U_a(x)$ and $f_{\rm HI}$, $\mathcal{F}_\mathrm{HI}(a)$ flattens out beyond $a\sim\mathrm{few}$, approaching its asymptotic value $\mathcal{F}_\mathrm{HI}$.

It is worth noting that these are the \textit{flux}-averaged atomic and molecular fractions, rather than the \textit{mass}-averaged fractions at a given radius. The density of material at a given radius is proportional to the flux divided by the velocity, so the \textit{mass}-averaged H~\textsc{i} fraction is given by
\begin{equation}
    \mathcal{F}_\mathrm{HI,mass}(a) = \frac{\int_{-\infty}^{x_\mathrm{crit}} f_\mathrm{HI}(a,x) p_M(x) / U_a(x)\, dx}{\int_{-\infty}^{x_\mathrm{crit}} p_M(x) / U_a(x) \, dx}.
    \label{eq:Fhi_mass}
\end{equation}
However, in practice this definition differs from the flux-averaged one only in marginal cases. For this reason we will for the rest of this paper focus exclusively on flux-averaged mass fractions as defined by \autoref{eq:Fhi_flux}.

\section{Results}
\label{sec:results}

The atomic and molecular fractions in the model we have just described are fully determined by our model for cloud expansion (i.e., clouds of constant area versus constant solid angle), the gravitational potential, and three dimensionless numbers: the Eddington ratio $\Gamma$, the Mach number $\mathcal{M}$ of the galaxy from which the wind is driven, and the ionisation parameter-like quantity $\xi$. In \autoref{ssec:parameters} we seek to understand how the dimensionless numbers affect the molecular fraction in winds for a fiducial choice of cloud area and potential shape, and in \autoref{ssec:application} we apply what we have learned to samples of observed galaxies.

\subsection{Which parameters determine the phase balance?}
\label{ssec:parameters}

\begin{figure}
	\includegraphics[width=\columnwidth]{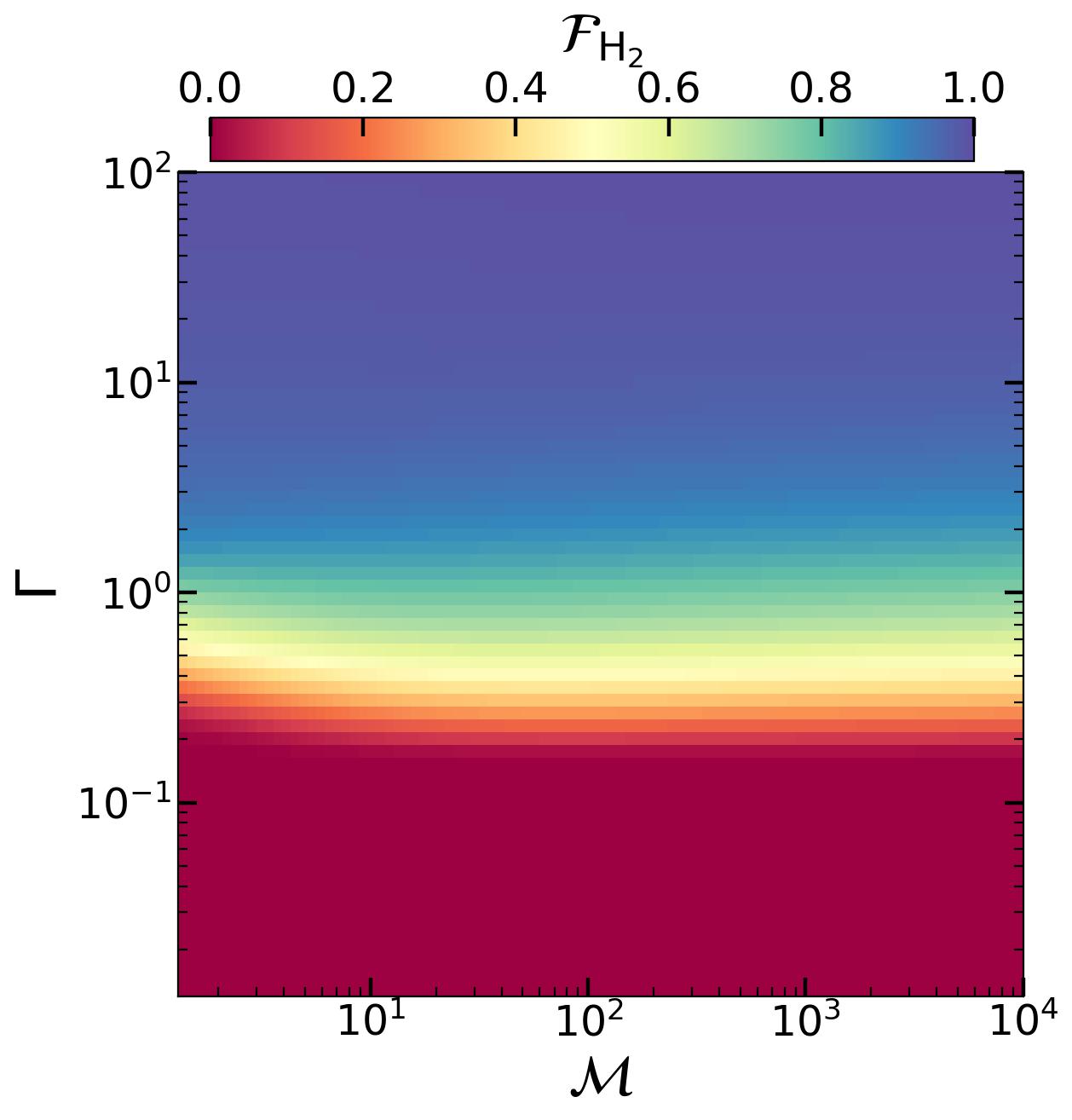}
 
    \caption{Flux-averaged asymptotic molecular fraction (\autoref{eq:Fhi_flux}) as a function of $\Gamma$ and $\mathcal{M}$ for fixed $\xi = 0.0406$ and clouds of fixed area in a point potential. We see that $\mathcal{F}_\mathrm{H_2}$ is nearly independent of $\mathcal{M}$ for $\mathcal{M}\gtrsim 10$, indicating it is not an important parameter determining how molecular the outflows are. }
    \label{fig:gamma_vs_mach}
\end{figure}

We are now in a position to determine how the dimensionless numbers we have identified affect the outflow chemical state. We begin this discussion with the parameters $\Gamma$ and $\mathcal{M}$; \autoref{fig:gamma_vs_mach} shows how $\mathcal{F}_\mathrm{H_2}$ varies with these parameters at fixed $\xi = 0.0406$ and for the case of constant area clouds in a point potential; we choose these example parameters based on the analysis of the wind of M82 by \citet{Yuan23a}, who find that the H~\textsc{i} and CO outflows are best described as fixed area, point potential, and where the value of $\xi$ comes from using their preferred galaxy parameters together with a gas surface density taken from \citet{Kennicutt+98a} in the formalism we describe below in \autoref{ssec:application}. However, the qualitative results are similar for any other parameter choice. In particular, we see that the asymptotic molecular fraction is nearly independent of the Mach number in the disc once $\mathcal{M} \gtrsim 10$, a condition that will be satisfied with respect to molecular gas in almost any galaxy. The physical origin of this behaviour is easy to understand if we examine \autoref{fig:uax_fhi}: at large $\mathcal{M}$, the column density PDF in the disc becomes very broad, so that a majority of the mass flux is carried by clouds with $x \ll x_\mathrm{crit}$ (redder colours in \autoref{fig:uax_fhi}). For these clouds gravity is unimportant ($\Gamma e^{-x} \gg 1$ in \autoref{eqn:uax}), and they quickly accelerate to a terminal velocity that is inversely proportional to their surface density. This means that the time they spend near the galaxy being exposed to strong dissociating radiation is directly proportional to their surface density -- but the amount of time required to fully dissociate a cloud is also linearly proportional to its surface density, and so this results in a molecular fraction that becomes independent of $x$ for $x\ll x_\mathrm{crit}$. This explains the behaviour shown in \autoref{fig:gamma_vs_mach}: once $\mathcal{M}$ is large enough that most flux is carried by clouds with $x \ll x_\mathrm{crit}$, further increases in $\mathcal{M}$ have no more effect on $\mathcal{F}_\mathrm{H_2}$. We may therefore discard $\mathcal{M}$ as an important parameter, and simply adopt the $\mathcal{M} \gg 1$ limit; for subsequent numerical results, we therefore set $\mathcal{M}=1000$.

\begin{figure}
\begin{center}

	\includegraphics[width=0.9\columnwidth]{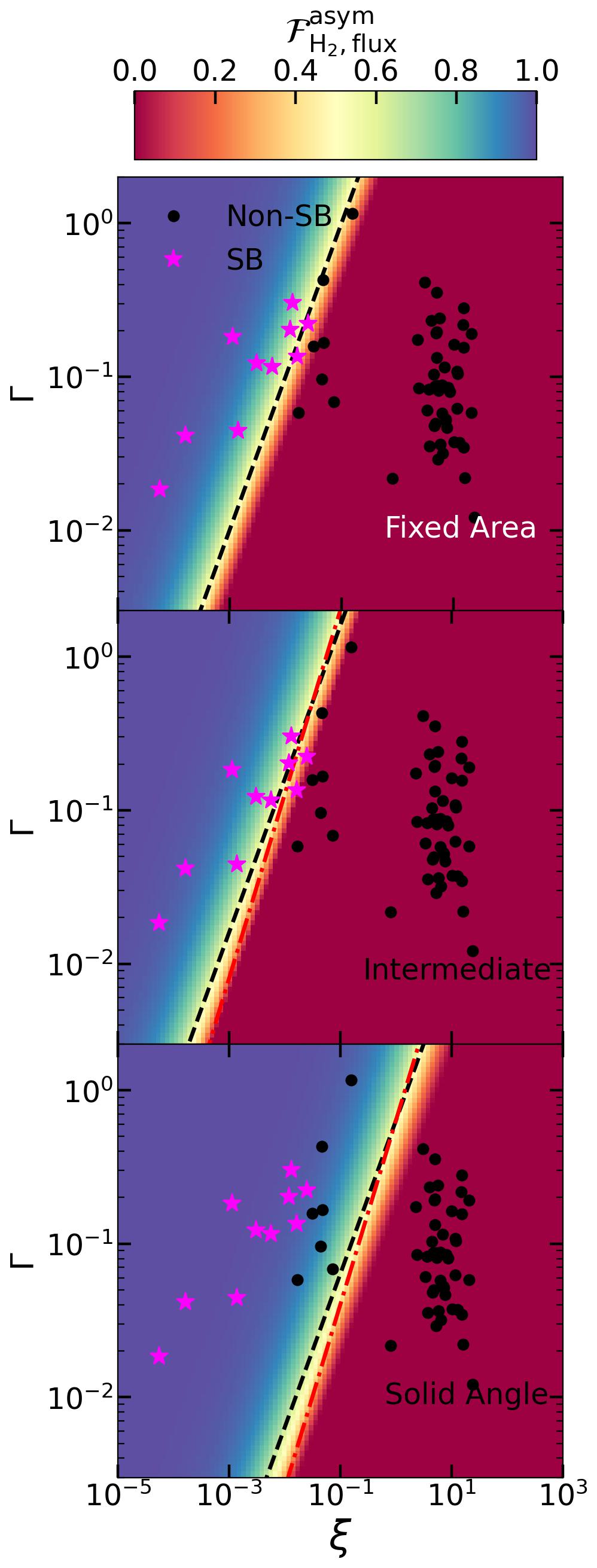}
    \caption{Same as \autoref{fig:gamma_vs_mach} except now we show the dissociation parameter $\xi$ on the horizontal axis. The different panels show different expansion laws -- fixed area (top), intermediate (middle) and constant solid angle (bottom). Black and magenta data points show the parameters of observed starburst and non-starburst galaxies, respectively; see \autoref{ssec:application} for details. Dashed and dot-dashed lines indicates a rough by-eye fit to the locus $\mathcal{F}_\mathrm{H_2} = 0.5$ for point potentials (black dashed) and isothermal potentials (red dot-dashed). Note that there is no line for fixed area clouds an in isothermal potential, since that case does not generate a wind that escapes.}
    \label{fig:gamma_vs_xi}
\end{center}
\end{figure}

Having discarded $\mathcal{M}$, we now turn to the two remaining parameters \Gam\ and \zai. We examine these two in \autoref{fig:gamma_vs_xi}; the colourmap in each panel is identical to the one shown in \autoref{fig:gamma_vs_mach} except that instead of $\mathcal{M}$ on the horizontal axis we have \zai. The three panels show different expansion laws -- fixed area (top), intermediate (middle) and the constant solid angle case (bottom). Evidently, the interplay between \Gam\ and \zai\ is the key in determining winds' molecular content. There is clear-cut separation between parts of parameter space where they are predominantly atomic and those where they are predominantly molecular, with the latter occupying the low $\xi$, high $\Gamma$ corner of parameter space. The dashed black lines in each panel show rough by-eye fits to the line separating the two regimes, which follow the functional form $\log \Gamma = k\log \xi + \mathrm{const}$ with slope $k=1$. Intuitively, galaxies to the right and below this line have stronger radiation fields and clouds that accelerate more slowly, meaning that the clouds spend a longer time in the presence of a radiation field, resulting in purely atomic outflows. Galaxies on the other side of the line have short dynamical times and rapidly-accelerating clouds, which propels clouds away from this radiation field faster, thereby preserving a high molecular fraction.

Comparing the three panels, we see that as cloud expansion becomes more rapid, the flow becomes increasingly molecular and the black dashed lines move right, albeit by a relatively small amount. The physical origin of this effect is that, even though expansion makes clouds bigger targets for dissociating photons and thus increases their dissociation rate, it also allows them to absorb momentum from the outflow more quickly and thus escape from the region of strong dissociation faster; the latter effect proves to be (slightly) more important. By contrast, the shape of the potential has minimal effects. The red dot-dashed lines in \autoref{fig:gamma_vs_xi} are identical to the black dashed ones, except for the case of an isothermal rather than a point-mass potential, and are clearly nearly the same. Potential shape is unimportant because it matters most once clouds have travelled some distance from the central galaxy, and by the time clouds get that far, per \autoref{fig:uax_fhi}, their chemical state is already set.

\subsection{Application to observed galaxies}
\label{ssec:application}

\begin{figure}
\begin{center}

	\includegraphics[width=0.8\columnwidth]{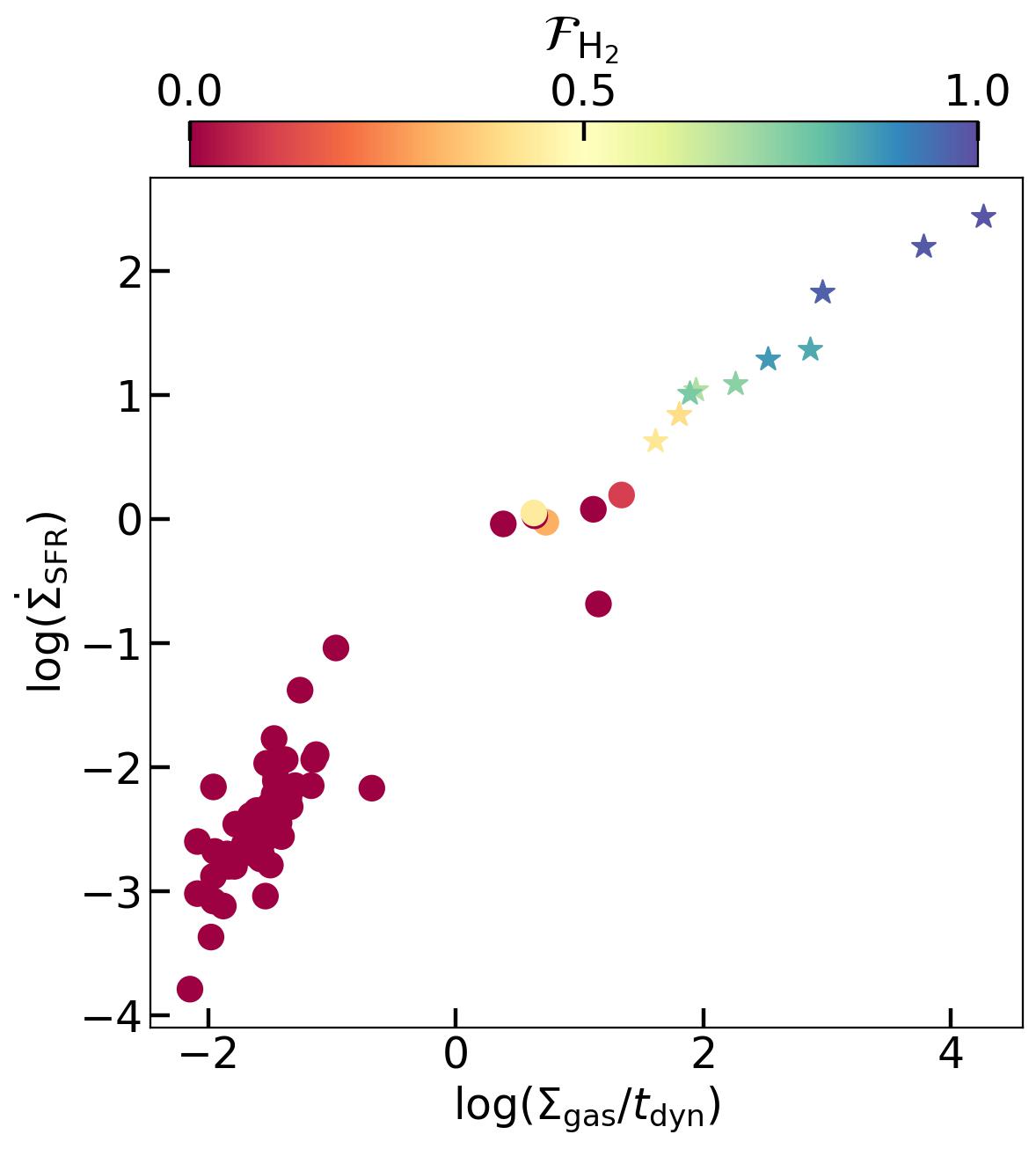}
    \caption{Star formation rate per unit area versus gas surface density per dynamical time for observed galaxies. Points are coloured by estimated outflow H$_2$ fraction $\mathcal{F}_\mathrm{H_2}$; circles show normal galaxies, stars show starbursts. See \autoref{ssec:application} for details on the data.
    }
    \label{fig:sigSFR_sigGas}
\end{center}
\end{figure}

Using the formalism described in \autoref{sec:the_model}, we can make predictions about the molecular/atomic fractions in outflows from observed galaxies. To do so, however, we require a method to estimate the dimensionless parameters $\Gamma$ and $\xi$ from observables. First consider the generalised Eddington ratio, which is related to galaxy properties as \citep{Thompson16a, Krumholz+17}
\begin{equation}
    \Gamma = \frac{\dot{p}}{4\pi G M_0 \overline{\Sigma}_0},
    \label{eq:Gamma}
\end{equation}
where $\dot{p}$ is the rate of momentum injection into the wind over all $4\pi$ sr and $M_0$ and $\overline{\Sigma}_0$ are the gravitating mass and gas surface density interior to the wind launch radius $r_0$. Since $M_0$ is not directly observable, we rewrite it in terms of the orbital period, which is. Specifically, the orbital speed is $v_\mathrm{orb} = \sqrt{G M_0/r_0}$ and the orbital period is $t_\mathrm{orb} = 2\pi r_0 / v_\mathrm{orb}$, and thus we have $M_0 = 4\pi^2 r_0^3/G t_\mathrm{orb}$. Similarly, the momentum injection is not known, but for a supernova-driven wind we can estimate that the terminal momentum available per unit mass of stars formed after cooling losses and neglecting any enhancement due to clustering is $\langle p_*/m_*\rangle \approx 3000$ km s$^{-1}$ \citep[e.g.,][]{Martizzi15a, Kim17a, Gentry19a}. Thus we can write $\dot{p} = \pi r_0^2 \dot{\Sigma}_* \langle p_*/m_*\rangle$, where $\dot{\Sigma}_*$ is the star formation rate per unit area. Inserting these expressions into \autoref{eq:Gamma} gives
\begin{equation}
    \Gamma = \frac{\dot{\Sigma}_* t_\mathrm{orb}^2}{16 \pi^2 r_0 \overline{\Sigma}_0}.
\end{equation}
The right hand side now depends only on the surface densities of gas and star formation, the size of the galaxy, and the orbital period, all quantities that are directly observable. Similarly, for the dissociation parameter $\xi$ (\autoref{eq:xi_def}), we take $t_0 = r_0 / v_0 = t_\mathrm{orb}/2^{3/2}\pi$. The most uncertain parameter is $\chi_0$, the dissociating photon flux in the galaxy; this could be either higher or lower than in the Milky Way for more intensely star-forming galaxies -- higher because the photon injection rate is higher due to more intense star formation, lower because there is also more gas and thus more extinction. We will adopt $\chi_0 = 1$ out of ignorance, and discuss the implications of this uncertainty below. Given these approximations, we have
\begin{equation}
    \xi = \frac{\dot{\Sigma}_\mathrm{diss,0} t_\mathrm{orb}}{2^{3/2}\pi \overline{\Sigma}_0},
\end{equation}
and again all quantities on the right hand side are now direct observables.

We take measurements of $\overline{\Sigma}_0$, $\dot{\Sigma}_*$, $r_0$, and $t_\mathrm{orb}$ from a subset of the galaxy sample compiled by \citet{Krumholz+12}, drawn in turn from \citet{Kennicutt+98a}, \citet{Daddi+10a}, and \citet{Genzel+10}. We take $\dot{\Sigma}_*$ and $\overline{\Sigma}_0$ directly from \citeauthor{Krumholz+12}'s tables; these have been standardised to use a common initial mass function and model for $\alpha_\mathrm{CO}$. For the galaxies in \citeauthor{Krumholz+12}'s compilation that originally come from \citeauthor{Daddi+10a}, we take radii and orbital periods from \citeauthor{Daddi+10a}'s tables. For those that come from \citeauthor{Genzel+10}, we take $r_0$ from their reported half-light radii, and we set $t_\mathrm{orb}=2\pi r_0/v_d$, where $v_d$ is their reported maximum circular velocity. Finally, for the subset of the \citeauthor{Krumholz+12} compilation drawn from \citeauthor{Kennicutt+98a}, we use \citeauthor{Kennicutt+98a}'s reported orbital periods, and we derive radii from the angular sizes reported in \citet{Kennicutt89} together with distances from SIMBAD \citep{Simbad}. 

We show the location of this galaxy sample in the $(\Gamma,\xi)$ plane in \autoref{fig:gamma_vs_xi}; points are colour-coded by whether \citet{Krumholz+12} classify each galaxy as a starburst or a non-starburst. We also show the data on the traditional Kennicutt-Schmidt plot colour-coded by derived $\mathcal{F}_\mathrm{H_2}$ (assuming constant area clouds) in \autoref{fig:sigSFR_sigGas}. The clear conclusion from these figures is that there is a transition from atomic-dominated outflows in non-starburst galaxies to molecular-dominated outflows in strong starbursts, with weak circumnuclear and dwarf starbursts (e.g., M82) and high-redshift giant star-forming discs straddling the boundary. Near this boundary the molecular fraction depends on cloud expansion law and on the poorly known dissociating radiation flux $\chi_0$ (to which $\xi$ is proportional), but most galaxies lie far enough from the boundary that neither of these uncertainties plausibly change the result. We note that our finding that outflows is galaxies like the Milky Way should completely dissociate by the time they reach distances of a few $r_0$, while those in weak starburst galaxies like M82 should retain a substantial molecular component to distances much larger than $r_0$, is in fact consistent with what is observed in those galaxies (Milky Way: \citealt{Di-Teodoro20a, Noon23a}; M82: \citealt{Leroy15b, Yuan23a}). We can further note that the starburst and non-starburst samples have fairly similar values of $\Gamma$ -- the higher momentum injection in the starbursts is largely balanced by their higher star formation rates -- and differ primarily in their values of $\xi$. This is driven in turn by the fact that starbursts have both high gas surface density and short dynamical times compared to normal galaxies.

\section{Discussion and Conclusion}
\label{sec:conclusion}

We present an analytic model to describe the molecular content of galactic winds. We first show that, unlike the balance between ionised and neutral material in winds, the balance between neutral and molecular hydrogen is almost certainly controlled by radiative rather than collisional processes, and thus the presence of molecules in winds that contain any amount of neutral material is determined by the galactic photon field. As a result, the molecular fraction is determined primarily by two dimensionless parameters: an ionisation parameter-like quantity $\xi$ that describes the ratio of the dissociation and dynamical times in the galaxy, and a generalised Eddington ratio $\Gamma$ that controls how quickly clouds in the wind are accelerated away from the disc and its dissociating radiation field. By contrast, other parameters such as the density distribution in the wind-launching galaxy (as parameterised in our model by the turbulent Mach number), the shape of the galactic potential, and the rate at which clouds that are swept into the wind expand all have only marginal effects.

Applying this model to a sample of observed galaxies, we predict a dichotomy between starbursts and normal star-forming galaxies, with the non-starbursts having predominantly atomic composition in the neutral wind phase and the starbursts predominantly molecular. The difference arises because starburst have both higher gas surface densities and shorter dynamical times than normal star-forming galaxies; the former factor means that dissociating radiation requires longer to eat through the molecular material, while the latter factor means that outflows escape quickly and thus there is less time available for the radiation to do so. 

Our predictions should be testable by future campaigns studying galactic outflows. In particular, as a new generation of H~\textsc{i} facilities that will be much more sensitive to low surface-brightness extended emission (e.g., MeerKAT, ASKAP, and eventually the full SKA) come online, it should be possible to identify a much larger number of atomic-dominated outlows than the handful currently known. Our model predicts that such outflows should be common for normal star-forming galaxies, but should decline in importance relative to molecular outflows as we move to the strong starburst regime. In particular, we predict that strong starbursts like Arp 220 should have substantial molecular outflows, but that these outflows should comparatively poor in atomic hydrogen.

\section*{Acknowledgements}

AV and MRK acknowledge support from the Australian Research Council through awards FL220100020 and DP230101055. The authors thank J.~Chisholm and D.~Fisher for helpful conversations.

\section*{Data Availability}

No new data were obtained as part of this work.



\bibliographystyle{mnras}
\bibliography{output.bbl} 

\begin{thebibliography}{}
\makeatletter
\relax
\def\mn@urlcharsother{\let\do\@makeother \do\$\do\&\do\#\do\^\do\_\do\%\do\~}
\def\mn@doi{\begingroup\mn@urlcharsother \@ifnextchar [ {\mn@doi@} {\mn@doi@[]}}
\def\mn@doi@[#1]#2{\def\@tempa{#1}\ifx\@tempa\@empty \href {http://dx.doi.org/#2} {doi:#2}\else \href {http://dx.doi.org/#2} {#1}\fi \endgroup}
\def\mn@eprint#1#2{\mn@eprint@#1:#2::\@nil}
\def\mn@eprint@arXiv#1{\href {http://arxiv.org/abs/#1} {{\tt arXiv:#1}}}
\def\mn@eprint@dblp#1{\href {http://dblp.uni-trier.de/rec/bibtex/#1.xml} {dblp:#1}}
\def\mn@eprint@#1:#2:#3:#4\@nil{\def\@tempa {#1}\def\@tempb {#2}\def\@tempc {#3}\ifx \@tempc \@empty \let \@tempc \@tempb \let \@tempb \@tempa \fi \ifx \@tempb \@empty \def\@tempb {arXiv}\fi \@ifundefined {mn@eprint@\@tempb}{\@tempb:\@tempc}{\expandafter \expandafter \csname mn@eprint@\@tempb\endcsname \expandafter{\@tempc}}}

\bibitem[\protect\citeauthoryear{{Bolatto} et~al.,}{{Bolatto} et~al.}{2013}]{Bolatto+13}
{Bolatto} A.~D.,  et~al., 2013, \mn@doi [\nat] {10.1038/nature12351}, \href {https://ui.adsabs.harvard.edu/abs/2013Natur.499..450B} {499, 450}

\bibitem[\protect\citeauthoryear{{Butsky}, {Hummels}, {Hopkins}, {Quinn}  \& {Werk}}{{Butsky} et~al.}{2024}]{Butsky24b}
{Butsky} I.~S.,  {Hummels} C.~B.,  {Hopkins} P.~F.,  {Quinn} T.~R.,   {Werk} J.~K.,  2024, \mn@doi [arXiv e-prints] {10.48550/arXiv.2402.03419}, \href {https://ui.adsabs.harvard.edu/abs/2024arXiv240203419B} {p. arXiv:2402.03419}

\bibitem[\protect\citeauthoryear{{Chen} \& {Oh}}{{Chen} \& {Oh}}{2023}]{Chen23a}
{Chen} Z.,  {Oh} S.~P.,  2023, \mn@doi [arXiv e-prints] {10.48550/arXiv.2311.04275}, \href {https://ui.adsabs.harvard.edu/abs/2023arXiv231104275C} {p. arXiv:2311.04275}

\bibitem[\protect\citeauthoryear{{Daddi} et~al.,}{{Daddi} et~al.}{2010}]{Daddi+10a}
{Daddi} E.,  et~al., 2010, \mn@doi [\apj] {10.1088/0004-637X/713/1/686}, \href {https://ui.adsabs.harvard.edu/abs/2010ApJ...713..686D} {713, 686}

\bibitem[\protect\citeauthoryear{{Di Teodoro}, {McClure-Griffiths}, {Lockman}, {Denbo}, {Endsley}, {Ford}  \& {Harrington}}{{Di Teodoro} et~al.}{2018}]{Di-Teodoro18a}
{Di Teodoro} E.~M.,  {McClure-Griffiths} N.~M.,  {Lockman} F.~J.,  {Denbo} S.~R.,  {Endsley} R.,  {Ford} H.~A.,   {Harrington} K.,  2018, \mn@doi [\apj] {10.3847/1538-4357/aaad6a}, \href {https://ui.adsabs.harvard.edu/abs/2018ApJ...855...33D} {855, 33}

\bibitem[\protect\citeauthoryear{{Di Teodoro}, {McClure-Griffiths}, {Lockman}  \& {Armillotta}}{{Di Teodoro} et~al.}{2020}]{Di-Teodoro20a}
{Di Teodoro} E.~M.,  {McClure-Griffiths} N.~M.,  {Lockman} F.~J.,   {Armillotta} L.,  2020, \mn@doi [\nat] {10.1038/s41586-020-2595-z}, \href {https://ui.adsabs.harvard.edu/abs/2020Natur.584..364D} {584, 364}

\bibitem[\protect\citeauthoryear{{Draine}}{{Draine}}{1978}]{Draine78a}
{Draine} B.~T.,  1978, \mn@doi [\apjs] {10.1086/190513}, \href {http://adsabs.harvard.edu/abs/1978ApJS...36..595D} {36, 595}

\bibitem[\protect\citeauthoryear{{Fielding} \& {Bryan}}{{Fielding} \& {Bryan}}{2022}]{Fielding22a}
{Fielding} D.~B.,  {Bryan} G.~L.,  2022, \mn@doi [\apj] {10.3847/1538-4357/ac2f41}, \href {https://ui.adsabs.harvard.edu/abs/2022ApJ...924...82F} {924, 82}

\bibitem[\protect\citeauthoryear{{Gentry}, {Krumholz}, {Madau}  \& {Lupi}}{{Gentry} et~al.}{2019}]{Gentry19a}
{Gentry} E.~S.,  {Krumholz} M.~R.,  {Madau} P.,   {Lupi} A.,  2019, \mn@doi [\mnras] {10.1093/mnras/sty3319}, \href {https://ui.adsabs.harvard.edu/\#abs/2019MNRAS.483.3647G} {483, 3647}

\bibitem[\protect\citeauthoryear{{Genzel} et~al.,}{{Genzel} et~al.}{2010}]{Genzel+10}
{Genzel} R.,  et~al., 2010, \mn@doi [\mnras] {10.1111/j.1365-2966.2010.16969.x}, \href {https://ui.adsabs.harvard.edu/abs/2010MNRAS.407.2091G} {407, 2091}

\bibitem[\protect\citeauthoryear{{Ginolfi} et~al.,}{{Ginolfi} et~al.}{2017}]{Ginolfi17a}
{Ginolfi} M.,  et~al., 2017, \mn@doi [\mnras] {10.1093/mnras/stx712}, \href {https://ui.adsabs.harvard.edu/abs/2017MNRAS.468.3468G} {468, 3468}

\bibitem[\protect\citeauthoryear{{Glover} \& {Mac Low}}{{Glover} \& {Mac Low}}{2007}]{Glover07a}
{Glover} S.~C.~O.,  {Mac Low} M.-M.,  2007, \mn@doi [\apjs] {10.1086/512238}, \href {http://adsabs.harvard.edu/abs/2007ApJS..169..239G} {169, 239}

\bibitem[\protect\citeauthoryear{{Gong}, {Ostriker}  \& {Wolfire}}{{Gong} et~al.}{2017}]{Gong17a}
{Gong} M.,  {Ostriker} E.~C.,   {Wolfire} M.~G.,  2017, \mn@doi [\apj] {10.3847/1538-4357/aa7561}, \href {http://adsabs.harvard.edu/abs/2017ApJ...843...38G} {843, 38}

\bibitem[\protect\citeauthoryear{{Huang}, {Katz}, {Dav{\'e}}, {Oppenheimer}, {Weinberg}, {Fardal}, {Kollmeier}  \& {Peeples}}{{Huang} et~al.}{2020}]{Huang20b}
{Huang} S.,  {Katz} N.,  {Dav{\'e}} R.,  {Oppenheimer} B.~D.,  {Weinberg} D.~H.,  {Fardal} M.,  {Kollmeier} J.~A.,   {Peeples} M.~S.,  2020, \mn@doi [\mnras] {10.1093/mnras/staa135}, \href {https://ui.adsabs.harvard.edu/abs/2020MNRAS.493....1H} {493, 1}

\bibitem[\protect\citeauthoryear{{Huang}, {Katz}, {Cottle}, {Scannapieco}, {Dav{\'e}}  \& {Weinberg}}{{Huang} et~al.}{2022}]{Huang22a}
{Huang} S.,  {Katz} N.,  {Cottle} J.,  {Scannapieco} E.,  {Dav{\'e}} R.,   {Weinberg} D.~H.,  2022, \mn@doi [\mnras] {10.1093/mnras/stab3363}, \href {https://ui.adsabs.harvard.edu/abs/2022MNRAS.509.6091H} {509, 6091}

\bibitem[\protect\citeauthoryear{{Indriolo} \& {McCall}}{{Indriolo} \& {McCall}}{2012}]{Indriolo12a}
{Indriolo} N.,  {McCall} B.~J.,  2012, \mn@doi [\apj] {10.1088/0004-637X/745/1/91}, \href {http://adsabs.harvard.edu/abs/2012ApJ...745...91I} {745, 91}

\bibitem[\protect\citeauthoryear{{Kennicutt}}{{Kennicutt}}{1989}]{Kennicutt89}
{Kennicutt} Robert~C. J.,  1989, \mn@doi [\apj] {10.1086/167834}, \href {https://ui.adsabs.harvard.edu/abs/1989ApJ...344..685K} {344, 685}

\bibitem[\protect\citeauthoryear{{Kennicutt}}{{Kennicutt}}{1998}]{Kennicutt+98a}
{Kennicutt} Robert~C. J.,  1998, \mn@doi [\apj] {10.1086/305588}, \href {https://ui.adsabs.harvard.edu/abs/1998ApJ...498..541K} {498, 541}

\bibitem[\protect\citeauthoryear{{Kim}, {Ostriker}  \& {Raileanu}}{{Kim} et~al.}{2017}]{Kim17a}
{Kim} C.-G.,  {Ostriker} E.~C.,   {Raileanu} R.,  2017, \mn@doi [\apj] {10.3847/1538-4357/834/1/25}, \href {http://adsabs.harvard.edu/abs/2017ApJ...834...25K} {834, 25}

\bibitem[\protect\citeauthoryear{{Kim} et~al.,}{{Kim} et~al.}{2020a}]{Kim+20}
{Kim} C.-G.,  et~al., 2020a, \mn@doi [\apj] {10.3847/1538-4357/aba962}, \href {https://ui.adsabs.harvard.edu/abs/2020ApJ...900...61K} {900, 61}

\bibitem[\protect\citeauthoryear{{Kim} et~al.,}{{Kim} et~al.}{2020b}]{Kim20b}
{Kim} C.-G.,  et~al., 2020b, \mn@doi [\apjl] {10.3847/2041-8213/abc252}, \href {https://ui.adsabs.harvard.edu/abs/2020ApJ...903L..34K} {903, L34}

\bibitem[\protect\citeauthoryear{{Krumholz}}{{Krumholz}}{2014}]{Krumholz14b}
{Krumholz} M.~R.,  2014, \mn@doi [\mnras] {10.1093/mnras/stt2000}, \href {http://adsabs.harvard.edu/abs/2014MNRAS.437.1662K} {437, 1662}

\bibitem[\protect\citeauthoryear{{Krumholz}, {McKee}  \& {Tumlinson}}{{Krumholz} et~al.}{2009}]{Krumholz+09}
{Krumholz} M.~R.,  {McKee} C.~F.,   {Tumlinson} J.,  2009, \mn@doi [\apj] {10.1088/0004-637X/693/1/216}, \href {https://ui.adsabs.harvard.edu/abs/2009ApJ...693..216K} {693, 216}

\bibitem[\protect\citeauthoryear{{Krumholz}, {Dekel}  \& {McKee}}{{Krumholz} et~al.}{2012}]{Krumholz+12}
{Krumholz} M.~R.,  {Dekel} A.,   {McKee} C.~F.,  2012, \mn@doi [\apj] {10.1088/0004-637X/745/1/69}, \href {https://ui.adsabs.harvard.edu/abs/2012ApJ...745...69K} {745, 69}

\bibitem[\protect\citeauthoryear{{Krumholz}, {Thompson}, {Ostriker}  \& {Martin}}{{Krumholz} et~al.}{2017}]{Krumholz+17}
{Krumholz} M.~R.,  {Thompson} T.~A.,  {Ostriker} E.~C.,   {Martin} C.~L.,  2017, \mn@doi [\mnras] {10.1093/mnras/stx1882}, \href {https://ui.adsabs.harvard.edu/abs/2017MNRAS.471.4061K} {471, 4061}

\bibitem[\protect\citeauthoryear{{Leroy} et~al.,}{{Leroy} et~al.}{2015}]{Leroy15b}
{Leroy} A.~K.,  et~al., 2015, \mn@doi [\apj] {10.1088/0004-637X/814/2/83}, \href {http://adsabs.harvard.edu/abs/2015ApJ...814...83L} {814, 83}

\bibitem[\protect\citeauthoryear{{Li} \& {Bryan}}{{Li} \& {Bryan}}{2020}]{Li20a}
{Li} M.,  {Bryan} G.~L.,  2020, \mn@doi [\apjl] {10.3847/2041-8213/ab7304}, \href {https://ui.adsabs.harvard.edu/abs/2020ApJ...890L..30L} {890, L30}

\bibitem[\protect\citeauthoryear{{Martini}, {Leroy}, {Mangum}, {Bolatto}, {Keating}, {Sandstrom}  \& {Walter}}{{Martini} et~al.}{2018}]{Martini18a}
{Martini} P.,  {Leroy} A.~K.,  {Mangum} J.~G.,  {Bolatto} A.,  {Keating} K.~M.,  {Sandstrom} K.,   {Walter} F.,  2018, \mn@doi [\apj] {10.3847/1538-4357/aab08e}, \href {https://ui.adsabs.harvard.edu/abs/2018ApJ...856...61M} {856, 61}

\bibitem[\protect\citeauthoryear{{Martizzi}, {Faucher-Gigu{\`e}re}  \& {Quataert}}{{Martizzi} et~al.}{2015}]{Martizzi15a}
{Martizzi} D.,  {Faucher-Gigu{\`e}re} C.-A.,   {Quataert} E.,  2015, \mn@doi [\mnras] {10.1093/mnras/stv562}, \href {http://adsabs.harvard.edu/abs/2015MNRAS.450..504M} {450, 504}

\bibitem[\protect\citeauthoryear{{Noon}, {Krumholz}, {Di Teodoro}, {McClure-Griffiths}, {Lockman}  \& {Armillotta}}{{Noon} et~al.}{2023}]{Noon23a}
{Noon} K.~A.,  {Krumholz} M.~R.,  {Di Teodoro} E.~M.,  {McClure-Griffiths} N.~M.,  {Lockman} F.~J.,   {Armillotta} L.,  2023, \mn@doi [\mnras] {10.1093/mnras/stad1890}, \href {https://ui.adsabs.harvard.edu/abs/2023MNRAS.524.1258N} {524, 1258}

\bibitem[\protect\citeauthoryear{{Raskutti}, {Ostriker}  \& {Skinner}}{{Raskutti} et~al.}{2017}]{Raskutti17a}
{Raskutti} S.,  {Ostriker} E.~C.,   {Skinner} M.~A.,  2017, \mn@doi [\apj] {10.3847/1538-4357/aa965e}, \href {http://adsabs.harvard.edu/abs/2017ApJ...850..112R} {850, 112}

\bibitem[\protect\citeauthoryear{{Rathjen} et~al.,}{{Rathjen} et~al.}{2021}]{Rathjen21a}
{Rathjen} T.-E.,  et~al., 2021, \mn@doi [\mnras] {10.1093/mnras/stab900}, \href {https://ui.adsabs.harvard.edu/abs/2021MNRAS.504.1039R} {504, 1039}

\bibitem[\protect\citeauthoryear{{Schneider}, {Ostriker}, {Robertson}  \& {Thompson}}{{Schneider} et~al.}{2020}]{Schneider20a}
{Schneider} E.~E.,  {Ostriker} E.~C.,  {Robertson} B.~E.,   {Thompson} T.~A.,  2020, \mn@doi [\apj] {10.3847/1538-4357/ab8ae8}, \href {https://ui.adsabs.harvard.edu/abs/2020ApJ...895...43S} {895, 43}

\bibitem[\protect\citeauthoryear{{Smith} et~al.,}{{Smith} et~al.}{2024}]{Smith24a}
{Smith} M.~C.,  et~al., 2024, \mn@doi [\mnras] {10.1093/mnras/stad3168}, \href {https://ui.adsabs.harvard.edu/abs/2024MNRAS.527.1216S} {527, 1216}

\bibitem[\protect\citeauthoryear{{Sternberg}, {Le Petit}, {Roueff}  \& {Le Bourlot}}{{Sternberg} et~al.}{2014}]{Sternberg14a}
{Sternberg} A.,  {Le Petit} F.,  {Roueff} E.,   {Le Bourlot} J.,  2014, \mn@doi [\apj] {10.1088/0004-637X/790/1/10}, \href {http://adsabs.harvard.edu/abs/2014ApJ...790...10S} {790, 10}

\bibitem[\protect\citeauthoryear{{Tacconi}, {Genzel}  \& {Sternberg}}{{Tacconi} et~al.}{2020}]{Tacconi20a}
{Tacconi} L.~J.,  {Genzel} R.,   {Sternberg} A.,  2020, \mn@doi [\araa] {10.1146/annurev-astro-082812-141034}, \href {https://ui.adsabs.harvard.edu/abs/2020ARA&A..58..157T} {58, 157}

\bibitem[\protect\citeauthoryear{{Tchernyshyov}}{{Tchernyshyov}}{2022}]{Tchernyshyov22a}
{Tchernyshyov} K.,  2022, \mn@doi [\apj] {10.3847/1538-4357/ac68e0}, \href {https://ui.adsabs.harvard.edu/abs/2022ApJ...931...78T} {931, 78}

\bibitem[\protect\citeauthoryear{{Thompson} \& {Krumholz}}{{Thompson} \& {Krumholz}}{2016}]{Thompson16a}
{Thompson} T.~A.,  {Krumholz} M.~R.,  2016, \mn@doi [\mnras] {10.1093/mnras/stv2331}, \href {http://adsabs.harvard.edu/abs/2016MNRAS.455..334T} {455, 334}

\bibitem[\protect\citeauthoryear{{Veilleux}, {Maiolino}, {Bolatto}  \& {Aalto}}{{Veilleux} et~al.}{2020}]{Veilleux20a}
{Veilleux} S.,  {Maiolino} R.,  {Bolatto} A.~D.,   {Aalto} S.,  2020, \mn@doi [\aapr] {10.1007/s00159-019-0121-9}, \href {https://ui.adsabs.harvard.edu/abs/2020A&ARv..28....2V} {28, 2}

\bibitem[\protect\citeauthoryear{{Vijayan}, {Krumholz}  \& {Wibking}}{{Vijayan} et~al.}{2024}]{Vijayan24a}
{Vijayan} A.,  {Krumholz} M.~R.,   {Wibking} B.~D.,  2024, \mn@doi [\mnras] {10.1093/mnras/stad3816}, \href {https://ui.adsabs.harvard.edu/abs/2024MNRAS.52710095V} {527, 10095}

\bibitem[\protect\citeauthoryear{{Walter} et~al.,}{{Walter} et~al.}{2017}]{Walter+17}
{Walter} F.,  et~al., 2017, \mn@doi [\apj] {10.3847/1538-4357/835/2/265}, \href {https://ui.adsabs.harvard.edu/abs/2017ApJ...835..265W} {835, 265}

\bibitem[\protect\citeauthoryear{{Weinberger} \& {Hernquist}}{{Weinberger} \& {Hernquist}}{2023}]{Weinberger23a}
{Weinberger} R.,  {Hernquist} L.,  2023, \mn@doi [\mnras] {10.1093/mnras/stac3708}, \href {https://ui.adsabs.harvard.edu/abs/2023MNRAS.519.3011W} {519, 3011}

\bibitem[\protect\citeauthoryear{{Wenger} et~al.,}{{Wenger} et~al.}{2000}]{Simbad}
{Wenger} M.,  et~al., 2000, \mn@doi [\aaps] {10.1051/aas:2000332}, \href {https://ui.adsabs.harvard.edu/abs/2000A&AS..143....9W} {143, 9}

\bibitem[\protect\citeauthoryear{{Yuan}, {Krumholz}  \& {Martin}}{{Yuan} et~al.}{2023}]{Yuan23a}
{Yuan} Y.,  {Krumholz} M.~R.,   {Martin} C.~L.,  2023, \mn@doi [\mnras] {10.1093/mnras/stac3241}, \href {https://ui.adsabs.harvard.edu/abs/2023MNRAS.518.4084Y} {518, 4084}

\makeatother
\end{thebibliography}





\bsp	
\label{lastpage}
\end{document}